\documentclass[runningheads]{llncs}

\usepackage{graphicx} 
\usepackage{xcolor}
\usepackage{xurl}
\usepackage{todonotes}
\usepackage{hyperref}

\usepackage{authblk}

\title{European Network For Gender Balance in Informatics (EUGAIN): Activities and Results}
\titlerunning{EUGAIN Activities and Results}

\author[1]{
Letizia Jaccheri, Barbora Buhnova, Birgit Penzenstadler, Karima Boudaoud, and Valentina Lenarduzzi}

\institute{Norwegian University of Science and Technology, Trondheim, Norway, \\ letizia.jaccheri@ntnu.no, \\
Masaryk University, Brno, Czech Republic, \\ buhnova@fi.muni.cz, \\
Chalmers University of Technology, Sweden, and Lappeenranta University of Technology, Finland, \\ birgitp@chalmers.se, \\
Université de Nice Sophia Antipolis, France, \\ Karima.BOUDAOUD@univ-cotedazur.fr, \\
University of Oulu, Finland, \\ Valentina.Lenarduzzi@oulu.fi}

\authorrunning{L. Jaccheri et al.}

\begin{document}

\maketitle

\section{Introduction }
This chapter provides a \textbf{summary of the activities and results of the European Network For Gender Balance in Informatics} (EUGAIN, EU COST Action CA19122). The main aim and objective of the network is to improve gender balance in informatics at all levels, from undergraduate and graduate studies to participation and leadership both in academia and industry, through the creation of a European network of colleagues working at the forefront of the efforts for gender balance in informatics in their countries and research communities.

Women are disproportionately represented in fields like Informatics (including Computer Science, Computer Engineering, Computing, ICT, software engineering) \cite{SEworldwide,SEeurope,silveira2019systematic}, spanning from undergraduate and graduate studies to leadership roles in academia and industry. Enhancing female participation in this domain presents a significant challenge for scholars, policymakers, and society as a whole as documented in several scientific studies \cite{glass2013s,bem1983gender,damon_handbook_2006,trinkenreich2022empirical,silveira2019systematic,ali2019discrimination,dastin2022amazon}. Despite widespread recognition of the issue, progress has been sluggish, despite ongoing efforts for change throughout Europe. The primary objective of this COST Action is to address the gender imbalance in Informatics by establishing and fortifying a diverse European network of academics actively advancing gender equality within their respective countries, institutions, and research communities. Leveraging their collective knowledge, experiences, challenges, successes, and failures, we aim to identify effective strategies that can be adapted and applied across various institutions and nations. Among its goals, the Action aims to provide the academic community, policymakers, industry stakeholders, and others with actionable recommendations and guidelines to tackle key challenges, including:

\begin{itemize}
    \item Increasing female enrollment in Informatics programs and careers; 
    \item Encouraging more female Ph.D. and postdoctoral researchers to pursue academic careers and apply for positions in Informatics departments
    \item Offering support and mentorship to empower young women in their professional journeys and address barriers preventing them from reaching senior roles.
\end{itemize}

\textbf{Chapter Structure}: Section \ref{sec:Background} describe the background, Section \ref{sec:Successful} describes the successful interventions while Section \ref{sec:Aim} the aim of this project. Section \ref{sec:Implementation} describes the project implementation. Section \ref{sec:Results} presents the obtained results. Section \ref{sec:Summary} summarizes the entire document.  

\section{Background}
\label{sec:Background}

When designing the European Network For Gender Balance in Informatics (EUGAIN), we started with a set of unstructured activities across Europe and knowledge derived from international research on the topic~\cite{jaccheri2020gender}. 

\subsection{Gender GAP in STEM}

The gender gap in STEM is widely discussed and recognized, but its relative size among various technology and engineering fields is less understood. Informatics (Computer Science, Computer Engineering, Computing, ICT) is one of the most heavily affected fields where the gender gap brings evident disparities. Areas such as Chemistry and Biology have significantly more balanced gender distribution (sometimes, the gender gap is even reversed, but only on lower career levels), whereas it is predominantly in Informatics, Engineering, and Technology that female absence is prevalent, with not much progress observed in the past years, whether in Europe~\cite{shefigures2021,InformaticsEurope} or the US~\cite{TaulbeeSurvey}. A study published when EUGAIN was set in June 2019, based on a comprehensive and up-to-date analysis of Computer Science literature, has estimated that the gender gap in Computer Science research (parity between the number of male and female authors) will not close for at least 100 years~\cite{Wang_2021}. 

\subsection{Informatics Higher Education in Europe}

Higher education statistics for European countries, collected over the past decade show that the strong female underrepresentation in Informatics higher education in Europe is a long-standing problem ~\cite{jaccheri2020gender}. 
At the Bachelor level, in Austria, Belgium, Denmark, Finland, Germany, Ireland, Italy, Latvia, Lithuania, the Netherlands, Poland, Spain, Switzerland, and the UK, 80\% or more of the students enrolling or graduating in Informatics Bachelor programs are male. In Bulgaria, Greece, Romania, and Estonia a slightly narrower gap exists, however, women do not represent more than 30\% of the Bachelor students~\cite{InformaticsEurope2}.
At the Master level participation of women increases in some countries, over 35\% of the Master graduates in Bulgaria, Romania and Greece, and around 30\% in the UK, Estonia, Ireland, and Latvia, but decreases in others, not surpassing 20\% of the Master graduates in Austria, Belgium, Czech Republic, Germany, Italy, Lithuania, the Netherlands, Poland, Portugal, Spain, Switzerland~\cite{InformaticsEurope2}. 
At the Ph.D. level, except for Bulgaria, Romania, Estonia, Turkey, all other countries have less than 25\% of women graduating from Informatics Ph.D. programs, corresponding in some cases to less than a handful of women, as the total number of Ph.D. graduates in many countries is quite small~\cite{shefigures2021,InformaticsEurope2}. 
A temporal analysis of the data shows that, on average, no significant progress in female participation in Informatics higher education has been observed over the past decade in Europe. The same is true for the US, as reported in~\cite{sax2017}

Participation in Computer Science was examined by Sax~\cite{sax2017}, gathering data on college students for four decades and highlighting a persistent, sizable under-representation of women in Computer Science in the US. 
Moreover, only a few women graduating with a Ph.D. in Informatics pursue an academic career, and even fewer progress to the highest academic ranks of an associate or a full professor. Similarly to other STEM areas, in Informatics the pipeline is leaking and glass ceiling persists. In the whole of Europe across all STEM where women and men are balanced in tertiary education, women still take less than 26\% of the full professor positions~\cite{shefigures2021}. The very low number of women reaching senior academic positions results in a scarcity of successful female role models to influence the new generations. To be a distinct minority in academia also results in the overload of invitations and requests (committees, administrative department roles, etc.), which penalizes women's careers, impacting negatively their research productivity, their work-life balance, their personal life, and health. 

\subsection{Gender Gap in the IT Industry}

The industry also inherits the male-dominated student population. Women are strongly underrepresented among ICT specialists in all EU Member States, which is in a striking contrast with total employment, where women and men are broadly balanced. Figures show that in 2021, an overwhelming majority (84.1\%) of ICT specialists employed in the EU were men~\cite{eurostatdata2021}. This was the case in every EU Member State, the highest shares of male ICT specialists were observed in the Czech Republic (92.6\%), Slovenia (90.8\%), France (89.7\%), Belgium (89.2\%), and Poland (89.1\%), while Bulgaria (63.4\%), Greece (70.6\%), Denmark (72.0\%) and Romania (70.8\%) recorded the lowest~\cite{eurostatdata}. The lack of women is among the reasons for the extensive skills and talent gap between the number of graduates in higher education institutions and the number of job positions available in ICT in Europe. Currently, an average of 53\% of European employers say they face difficulties in finding the right people with the right qualifications. The highest percentages were recorded in the Czech Republic (79\%), Austria (78\%), Malta (73\%), Luxembourg (71\%) the Netherlands (69\%), Slovenia (65\%), Germany (64\%) and Denmark (61\%)~\cite{eurostatdata}. Hundreds of thousands of vacancies for ICT professionals in Europe remain unfilled, and this gap grows as our society moves to a pervasively digitalized world built on unprecedented technological developments. The talent gap in ICT is one of the most serious threats to the economic development of Europe. 

The tech sector’s dominantly male workforce intrinsically promotes the creation and development of systems prone to gender bias. From smartphone voice assistants (Android’s Cortana, Apple’s Siri, Microsoft’s Alexa) that are all female with noticeable submissive personalities and ill-equipped to respond to user requests regarding crises that predominantly affect women (e.g., sexual assault)~\cite{Idbush}, to activity trackers that fail to measure steps in the, predominantly female activity of pushing a stroller. Transport networks that ignore the so-called “mobility of care” and AI recruiting technology developed trained predominantly on men’s résumés are among the many examples, more are found on the EU Report of the Expert Group “Innovation through Gender” and the website of the international project on Gendered Innovations. 
Despite the clear negative impact and consequences of a strongly gender-unbalanced environment, unfortunately, the fight for gender balance and equality in Informatics is seen as a women’s problem. Projects, programs, actions, and strategies are invariably led by highly motivated and achieving women who volunteer their time to establish a more equal environment and pave the way for the new generation of female scientists. Going beyond their daily work, they are responsible for the monumental effort, and comparatively more moderated funding, that has been spent on the efforts for gender equality in Informatics.

\section{Successful Interventions}
\label{sec:Successful}

Despite some overall discouraging numbers~\cite{InformaticsEurope2}, some remarkably successful examples at the university level are found in the US as well as in Europe. On a global level, we find work by UNESCO~\cite{Idbush}.

In the USA, the most famous examples are \href{https://www.hmc.edu/about/2018/05/15/harvey-mudd-graduates-highest-ever-percentage-of-women-physics-and-computer-science-majors/}{Harvey} Mudd College and \href{https://www.cmu.edu/news/stories/archives/2016/september/undergrad-women-engineering-computer-science.html}{Carnegie} Mellon where in the past decade gender parity has been achieved in Computer Science entrants and graduates~\cite{frieze2019computer}. Although inspiring, these efforts remain isolated and proved difficult to escalate to more institutions and to improve the national statistics. 

Europe still lags behind the US in regards to the amount of funding, successful examples, and the level of organization of the community. Organizations and groups such as AnitaB.org\cite{AnitaB}, ACM-W \cite{womenacmorg}, CRA Women \cite{CRA-WP}, \href{https://ncwit.org}{National}  Center for Women \& Information Technology, IEEE Women in Computing Committee \cite{women-in-computing}, Association for Women in Computing \cite{awc-hq}, and \href{https://girlswhocode.com}{Girls Who Code} in partnership with industry have established a thriving community empowered to inspire and encourage the new generations and to support the careers women in Computer Science. The most spectacular example of this community is the \href{https://ghc.anitab.org}{Grace Hopper Celebration}, which in 2023 gathered over 30,000 attendees from over 80 countries, almost all women, at all stages in Computer Science studies and careers, providing an invaluable opportunity for women to find inspiration, networking, and strategies to thrive in their careers. 

In Europe, we find examples of EU public-funded projects such as \href{https://equal-ist.eu}{EQUAL-IST}, \href{https://women4it.eu}{Women4IT}, and pan-European networks such as the \href{https://www.informatics-europe.org/society/women-in-icst-research-and-education/working-group.html}{Informatics Europe’s} Women in Informatics Research and Education Working Group and the \href{https://acmweurope.acm.org}{ACM-WE}  Committee (both more oriented to women in the academic career); the \href{https://cepis.org/women-in-ict/}{CEPIS} Women in ICT Task Force and the \href{https://ecwt.eu}{European Centre} for Women and Technology  (both more oriented to women in the ICT profession). Nevertheless, several commendable projects, internal policies, and strategies are found in many Universities, funded by national mechanisms, specifically to increase the number and retention of female students in Computer Science programs. Here are a few good examples: 
  \begin{itemize}
      \item The Bamberg  CS30  Strategy  \cite{uni-bamberg},  Faculty  of  Information Systems and Applied Computer Sciences, University of Bamberg, Germany -- Started in 2005 and aim at reaching a female/male ratio of at least 30\% across all Computer Science programs The number of women enrolling in first-year Computer Science studies has been increasing since 2013 and reached 37\% in 2017, establishing a new record in Germany.
      \item The Girl Project Ada \cite{ada} \cite{lagesen2022inclusion}, Faculty of Information Technology and Electrical Engineering, NTNU, Norway -- Started in 1997 and aims at recruiting more girls to the ICT studies and prevent dropouts. The female share of entrants in ICT studies has, on average across different programs, almost doubled, going over 25\% in the Computer Science program in 2017.
      \item CS4All initiative, School of Computer Science, TU Dublin, Ireland -- Started in 2012 and aims at increasing the number of female students coming to Computer Science undergraduate programs and reduce the numbers failing to progress in the critical first year. The female share of enrolled students in a new Computer Science Bachelor Program, with a strong emphasis on Internationalisation and Globalisation (22\%) is double the one of the standard Computer Science in the same period. Retention has been strongly improved, particularly for first-year students with an average 89\% progression from year 1 to year 2 (the most critical), now the highest progression rate for Computer Science in Ireland.
  \end{itemize}

Projects, internal policy, and strategy management for supporting the transition of female Ph.D. and Postdoctoral Researchers into Faculty positions and for developing the careers of female Faculty in Informatics Departments are also found across Europe, but the impact has been less significant, and the numbers of female researchers and professors, in general, remain discouragingly low. 

Here are a few institutions that have implemented beneficial internal strategies and policies to increase the number of female researchers and faculty and support their careers: 
 \begin{itemize}
     \item \href{https://informatics.tuwien.ac.at/women-in-informatics/}{Faculty of Informatics} TU Vienna, Austria
     \item \href{https://irishtechnews.ie/female-intake-computer-science-degrees-tu-dublin}{School of Computer Science}, TU Dublin, Ireland 
     \item \href{https://www.ul.ie/news/ul-backing-new-project-offering-pathway-for-young-women-to-computer-science-career}{Department of} Computer Science and Information Systems, University of Limerick, Ireland
     \item \href{https://women.cs.ru.nl}{Institute}   for  Computing  and  Information Sciences,  Radboud  University  in  Nijmegen,  the  Netherlands 
     \item \href{https://www.ntnu.edu/idun}{Faculty} of Information Technology and Electrical Engineering, NTNU, Norway 
     \item \href{https://www.qub.ac.uk/schools/eeecs/Connect/Equality-Diversity-and-Inclusion/WomensLeadershipProgramme}{School} of Electronics, Electrical Engineering \& Computer Science, Queen's University Belfast, UK 
     \item \href{https://www.information-age.com/uk-universities-leading-way-computer-science-gender-diversity-17429/}{Department} of Computer Science, University College of London, UK 
     \item \href{https://www.ed.ac.uk/informatics/news-events/news/2022/women-and-girls-in-informatics}{School} of Informatics, University of Edinburgh, UK 
     \item Department of Computer Science, University of Sheffield, UK \newline \url{https://www.sheffield.ac.uk/dcs/about/women-computer-science}
     \item \href{https://www.hs-bremen.de/en/study/degree-programme/international-womens-degree-programme-in-computer-science-bsc/}{Faculty} of Mathematics and Computer Science, University of Bremen, Germany 
     \item \href{https://www.cs.cit.tum.de/en/cs/department/diversity/wics/}{Department of Informatics, TU Munich, Germany} 
     \item Department of Informatics, University of Lille, France (initiative \href{https://edmadis.univ-lille.fr/en/phd-at-lille/women-and-digital-sciences}{1} and \href{https://www.ins2i.cnrs.fr/en/codebreakhers-digital-world}{2})
     \item \href{https://www.grenoble-inp.fr/en/about/grenoble-inp-for-gender-equality}{Grenoble INP, University Grenoble Alpes}, France
   
 \end{itemize}
 
Many more Universities have individual projects or fellowships (involving directly or indirectly Informatics Departments) aiming at improving gender balance. A few bold examples, involving substantial funding, include the \href{https://www.tue.nl/en/working-at-tue/scientific-staff/irene-curie-fellowship/}{Irène}  Curie Fellowship at TU Eindhoven, the \href{https://www.chalmers.se/om-chalmers/organisation-och-styrning/jamstalldhet/genie-gender-initiative-for-excellence/}{Gender}  Initiative for Excellence at Chalmers University of Technology, and the \href{https://www.ntnu.edu/idun}{IDUN project at NTNU}. Some National Informatics Associations and National Research Labs also have special interest groups, or Equal Opportunities offices with a focus on gender balance, to cite a few: \href{https://gi.de}{Gesellschaft für Informatik} in Germany, \href{https://ict-research.nl}{IPN} (ICT Research Platform Netherlands), \href{https://www.societe-informatique-de-france.fr}{Société Informatique de France} Inria, \href{https://www.mpi-inf.mpg.de/home/}{Max} Planck Institute for Informatics.

Moreover, over the course of EUGAIN, we have established the \href{https://www.informatics-europe.org/society/minerva-informatics-equality-award.html}{Minerva Awards}  within Informatics Europe and successfully presented the awards for 9 years in a row, usually at the European Computer Science Summit (ECSS).

\section{The Aims of EUGAIN}
\label{sec:Aim}



The overarching \textbf{challenge} of EUGAIN is to enhance the representation of women in the field of Informatics at various educational and professional levels. This involves strategies such as increasing the number of female students opting for Informatics in higher education, fostering an environment that ensures the retention and successful completion of studies by female students, and encouraging the participation of women in advanced academic roles, including Ph.D. and postdoctoral research positions. Additionally, the focus extends to supporting and inspiring young women in their careers, addressing key obstacles that hinder their progress toward senior positions within the field. To achieve these objectives, collaboration with network partners is essential, leveraging their experiences to overcome challenges and implement effective measures across diverse institutions and countries, leading to sustained positive outcomes in the long run.

The main \textbf{research questions} investigate actions, policies, measurable results, geographical perspective, relation to industry, relation to school, intersectionality, and role of male colleagues. Table \ref{tablequestions} provides research questions that were used when setting up investigations in this field of gender and computer science within EUGAIN. 

\begin{table}
\begin{center}
\small
\begin{tabular}{|p{\textwidth}|}
\hline
\textbf{Questions} \\
\hline
How successful have the implemented actions and policies been? How can their impact be measured? \\
\hline
How much effort (people and time) and funding were spent on projects that have had successful, measurable results? \\
\hline
How visible have these actions at the university, region, country level, or internationally been? \\
\hline
How to replicate successful actions and policies in different institutions or countries? \\
\hline
How many countries have pre-established national networks with a focus on gender balance in Informatics/ICT? Have these networks had a positive impact on results and outcomes? \\
\hline
What is the proportion of Informatics Departments across Europe that have never implemented any measure or policy to improve gender balance? \\
\hline
Has the industry been involved in these efforts? Has this type of collaboration influenced positively the results? If not, how to foster more effective and successful partnerships? \\
\hline
Have male colleagues been leading or actively participating in the projects and actions, or does this remain primarily a women’s problem? What has been different in cases with significant male engagement? Would the participation of more men (particularly in leading positions) have a positive impact on progress and results? \\
\hline
Why do some countries have better female participation in Informatics (studies or profession)? Why are they a minority? Are there cultural, historical, or economic reasons for this? \\
\hline
Are there Departments (or countries) that have policies for improving more general diversity and include other minorities? \\
\hline
Has the lack of Informatics as a foundational discipline in schools played an important role in the low numbers of female students in Informatics higher education? \\
\hline
\end{tabular}
\end{center}
\caption{Research questions used as starting point for investigations.}
\label{tablequestions}
\end{table}

\subsection{Objectives}

The objectives of EUGAIN are divided into two main categories: Research Coordination Objectives (Table \ref{ResearchCoordinationTable}) and Capacity-building Objectives (Table \ref{CapacityBuildingTable}).

\begin{table}
\begin{center}
\small
\begin{tabular}{|p{\textwidth}|}\hline
\textbf{Research Coordination Objectives} \\
\hline
Coordinate information gathering and collecting practices and initiatives for recruiting and retaining female students, researchers, and professors \\
    \hline
 Support partners in assessing and evaluating existing practices and methodologies, facilitating the choice of what could be implemented according to the local situation (cultures, resources, etc.); \\
    \hline
   Coordinate data  from  across  each  WG  to  assist  in  the  development of  cross-validated instruments to help Informatics Departments set goals and priorities for female recruiting, integration, and promotion; \\
    \hline
Collate collaboratively a handbook of interventions and web-based resources across all WGs for practical use by the academic community and stakeholders; \\
    \hline
 Deliver  guidance  and  recommendations  on   how  to  overcome  the  challenges  in   a comprehensive policy document targeting policymakers and other relevant stakeholders at the national and the EU level; \\
    \hline
Create visibility, both within the academic community and to other stakeholders, about the common issues and challenges facing the academic community, and local, national, and EU authorities in addressing gender balance in Informatics; \\
    \hline
Create a communications strategy to spread information about the actions and results to the general public and stakeholders, using a website created for the project, social media channels, newsletters, and press releases; \\
    \hline
 Develop a common European understanding around the issues of female participation in Informatics, policy priorities, and areas of intervention; Involve industry stakeholders in the efforts to address the main challenges and create opportunities and synergies; \\
    \hline
  Develop and publish an Action website to become the reference point for addressing gender balance in Informatics, including an online repository of the evidence collected by the WGs, information about the networking activities organized, and channels for dissemination and communication. \\
      \hline
\end{tabular}
\end{center}
\caption{Research coordination objectives.}
\label{ResearchCoordinationTable}
\end{table}

\begin{table}
\begin{center}
\small
\begin{tabular}{|p{\textwidth}|}\hline
\textbf{Capacity Building Objectives} \\
\hline
    \item Establish an efficient and lasting network of excellence to advance knowledge and methods to improve gender balance in Informatics;\\
      \hline
    \item Such a network shall encourage sustainable collaboration, and facilitate knowledge and experience sharing, with an emphasis on intervention best practices, through seminars, workshops, and short-term exchange visits, involving a comprehensive list of stakeholders;\\
      \hline
    \item Promote policy and intervention practices for recruiting and selection of female students, researchers, and professors, including guidelines for their monitoring and evaluation; disseminate the practices for further development by the wider academic community, fostering collaborative international projects;\\
      \hline
    \item Encourage publication (and support the drafting) of peer-reviewed papers, and presentations at important conferences and events, to create at an international level awareness of the gender gap in Informatics;\\
      \hline
    \item Increase awareness of the issues across disciplinary boundaries, both within and outside of academia, by promoting continued exchange and development of knowledge, practice, and policy guidance;\\
      \hline
    \item Cooperate with industry to foster career networks, creating mutually beneficial synergies, for students and early career researchers to find excellent career opportunities, and for the industry to tap into a pool of highly motivated talented individuals;\\
      \hline
    \item Act as a transnational platform facilitating multi-stakeholder engagement and co-creating processes and actions at local, national, European, and international levels.\\
        \hline
\end{tabular}
\end{center}
\caption{Capacity Building Objectives.}
\label{CapacityBuildingTable}
\end{table}

\section{Implementation through the Working groups}
\label{sec:Implementation}


To ensure progress beyond the state of the art and encourage novel approaches and methods, three Working Groups (WGs) addressing the challenges of each transition: from School to University (WG1); from Bachelor/Master studies to Ph.D (WG2); from Ph.D./Postdoc to Professor (WG3) have been established, combining experts and perspectives from different institutions and COST countries. Two additional WGs, on Cooperation with Industry and Society (WG4), and Strategy  \& Dissemination (WG5) support and promote outreach of the activities and outcomes. Other dissemination and communication activities ensure reaching all interested stakeholders. Moreover, tangible deliverables will also promote the advancement of the state of the art.  

\begin{table}
\begin{center}
\small
\begin{tabular}{|p{\textwidth}|}
\hline
\textbf{Main Deliverables} \\
\hline
A website (https://eugain.eu/); \\
\hline
a repository of initiatives and best practices;\\
\hline
booklets with practical recommendations (see section 6.1); \\
\hline
a handbook with validated measures and guidelines helping university departments to recruit and retain female (students, PhD students, professors and researchers);\\
\hline
policy recommendation documents for local, national and international institutions; \\
\hline
publications, and presentations (see section 6.2).\\
\hline

\end{tabular}
\end{center}
\caption{Deliverables}
\label{Deliverables}
\end{table}

The main goals and activities of the different working groups are described below.
\subsection{WG1: From School to University}
The main objective of WG1 was to update and design a new set of measures on how to promote the education and participation of more female students in Informatics higher education. Moreover, it aimed to increase the number of applications and to ensure that students who started will thrive, make their voices heard, and complete their studies. 
In terms of tasks and activities, WG1 focused on collecting and evaluating current initiatives existing in the COST countries and institutions part of the Action, including targeted recruitment initiatives, activities for students (from primary to high school), mentoring and career programmes in academia and industry. It also collated examples of how female students voice can be encouraged across universities in generating information and ideas.

\subsection{WG2: From Bachelor/Master Studies to Ph.D}
WG2 aimed to design a new set of measures on how to promote the participation of more female students in Ph.D. programs in Informatics and ensure that students who started will complete their Ph.D. studies. 
The main tasks of WG2 were to: (1) collect and assess cross-national action plans/guidelines (national or regional) to inform about research activities and role models in research and education; (2) collate current interventions/tools to inform about actions both in general terms and specifically regarding gender and diversity issues; (3) collate examples of how female Ph.D. student voices’ can be encouraged across universities in generating innovative research projects and ideas and (4) gather evidence on their effectiveness across different groups and with regards to gender and age systematically reviewing completeness of the information, degree of usage, local evaluations carried out, and sustainability. 

\subsection{WG3: From Ph.D. to Professor}
The main goal of WG3 was to identify successful practices to recruit more female professors in Informatics and to limit the dropout rate of women along the path to professorship and leader positions in academia. It aimed also to help to increase the proportion of women in international research projects. 
To reach this goal, WG3 focused on (1) collecting experiences from ongoing initiatives in COST countries universities and assess evidence \cite{rubegni2023owning}; (2) identifying HR policies and recruitment strategies aimed at increasing female recruitment and retention within Departments, Institutes/Faculties/Schools, Universities; (3) designing protocols for collaboration between the management and the employees at the faculty, with a focus on gender equality; (4) designing career development programme for Ph.D. students and postdoctoral researchers; (5) developing a mentor scheme for women at the master's level to associate professor level; (6) creating international mentoring schemes between women in scientific positions at different levels and in different COST countries and(7) developing a strategy for recruiting women in externally funded projects, especially for EU funding. 
\subsection{WG4: Cooperation with Industry and Society}
The main objective of WG4 aimed to assure that cooperation with stakeholders in industry and other sectors exists at a local, regional, national and EU level and that particular issues existing in each country are taken into consideration. It aimed also to analyse the existing practices put in place for university departments, institutes/faculties/schools to deal with external cooperation with a focus on gender issues and evaluate what assessment exists for these practices. 
The main tasks and activities of WG4 focused on: (1) collate evidence of successful industry-university collaboration across partners and countries \cite{razavian2015feminine,happe2021frustrations}; (2) gather and assess evidence of best practices on how collaboration with industry and other sectors have had positive impact on gender balance in Informatics/ICT; (3) collate action plans/guidelines on integration from national and regional authorities for policy evaluation and (4) engage with the IT/ICT sector to improve the integration of gender balance in their research portfolio and recruitment strategy. 

\subsection{WG5: Strategy \& Dissemination}
The objectives of WG5 were to: (1) raise awareness about the gender imbalance and bias in Informatics; (2) advocate and lobby for change; (3) disseminate the action results to all partners and national networks and (4) reach out to all external stakeholders.
To reach these objectives, the main task of WG5 was to assure that the main activities, events, outcomes and deliverables of all WGs have the most optimal visibility and reach the relevant stakeholders. This is done through the organization of an Annual European Workshop on gender balance in Informatics/ICT (during the project duration and on the longer term annually, after the end of the project) and face to face meetings with relevant policy officers at the EU level and national level (involving then the partner(s) in their country). Finally, WG5 is in charge of organizing an European Award for best practices in departments/institutes/schools/faculties of European universities and research labs that encourage and support the careers of women in Informatics research and education (selected by a review panel of international experts).


\section{Results and Outputs}
\label{sec:Results}
In this Section, we reported the main results we got during these four years as booklets and scientific outputs. 

\subsection{Booklets}
\label{sec:booklets}

We produced 4 booklets, a policy recommendation document, and a handbook of intervention methods, as follows:

\begin{itemize}
    \item \textit{Booklet ``From Ph.D. to Professor''}: that includes the best practices for supporting the transition of Ph.D. and postdoctoral researchers into faculty positions.
    \item \textit{Booklet ``From School to University'':} that includes the best practices and suggestions for recruiting and retaining female students.
    \item \textit{Booklet(s) ``Future Informatics Students'':} that includes advice and advantages of studying and choosing Informatics as a career.
    \item  \textit{Booklet ``From Bachelor/Master Studies to Ph.D.'':} that includes the best practices and suggestions for retaining and supporting the transition of female students to Ph.D. positions.
    \item \textit{Policy recommendation document:} that includes a set of policy recommendations directed to policymakers, at the national and European level. 
    \item \textit{Handbook of intervention methods:} that provides an understanding of the factors that contribute to increasing the recruitment and retention of female computer scientists, methods, and intervention strategies.     
\end{itemize}

\subsection{Scientific Output}

Based on the results obtained in this project, we published more than 50 papers both in conferences and journals. The complete list of publications is available in an online appendix \footnote{https://shorturl.at/abfzN}.

\section{Summary and Future Outlooks}
\label{sec:Summary}

By now we have understood the barriers and effective strategies towards improving gender balance in informatics, summarized also within the booklets discussed in Section \ref{sec:booklets}. EUGAIN has offered a platform for inspiration crossing cultural boundaries and for gathering insights on effective strategies towards addressing stereotypes, promoting role models, closing the confidence gap, growing a sense of belonging, and learning to give recognition and credit to all the talented women and other underrepresented talents in informatics.

At the same time, we now understand the importance of continuing this essential work. We see the immense importance of a better understanding of cultural differences and their influences and strategies that give us a better ability to engage and recognize diverse talent, together with effective tools to guide women and girls throughout the maze of educational and career decisions in the growing world of informatics and technology.

When starting EUGAIN, we began with an ambitious set of questions we intended to answer (see Table~\ref{tablequestions}). Now, four years later, while we have shed light on all these questions, we find them far from answered. We have understood how hard it is to measure progress and how multifaceted the progress can be, moving us into technology that is inclusive for everyone, not only women and girls. 

Over the duration of EUGAIN, we have had the privilege to watch closely when the change was taking place in the institutions of our project members, which we have celebrated with the Minerva award and made sure to document as many of the efforts take time to bloom. As for now, we are excited that the seeds have been planted and will continue nourishing these ongoing activities in all our institutions. We hope our results and this book bring inspiration to the reader to do the same.




\bibliographystyle{splncs04}
\bibliography{main}

\end{document}